\documentstyle[emulateapj, graphicx]{article}

\def\myputfigure#1#2#3#4#5%
{\hskip0.03\textwidth\vskip#5pt
\makebox[0pt]{\hskip#2in
\includegraphics[width=#3\textwidth]{#1}}\vskip#4pt\hfill}
\makeatletter

\newenvironment{tablehere}
  {\def\@captype{table}}
  {}
\makeatother

\def\fun#1#2{\lower3.6pt\vbox{\baselineskip0pt\lineskip.9pt
\ialign{$\mathsurround=0pt#1\hfil##\hfil$\crcr#2\crcr\sim\crcr}}}
\def\lap{\mathrel{\mathpalette\fun <}}

\def\mass{{\cal M}}

\def\msun{{\mass_\odot}}

\def\beq{\begin{equation}}
\def\eeq{\end{equation}}
\def\mh{\mbox{M}_{bh}}

\lefthead{Triaxial Nuclei}
\righthead{}

\begin{document}

\title{Triaxial Black-Hole Nuclei}

\author{M. Y. Poon \& D. Merritt}

\affil{Department of Physics and Astronomy, Rutgers University,
New Brunswick, NJ 08855}

\begin{abstract}
We demonstrate that the nuclei of galaxies containing supermassive
black holes can be triaxial in shape.
Schwarzschild's method was first used to construct self-consistent
orbital superpositions representing nuclei with axis ratios of 
$1:0.79:0.5$ and containing a central point mass representing a black hole.
Two different density laws were considered,
$\rho\propto r^{-\gamma}$, $\gamma=\{1,2\}$. 
We constructed two solutions for each $\gamma$: one containing only
regular orbits and the other containing both regular and chaotic orbits.
Monte-Carlo realizations of the models were then 
advanced in time using an $N$-body code to verify their stability.
All four models were found to retain their triaxial shapes for many crossing
times.
The possibility that galactic nuclei may be triaxial complicates the 
interpretation of stellar-kinematical data from the centers of galaxies
and may alter the inferred interaction rates between stars and supermassive
black holes.

\end{abstract}


\section {Introduction}
The phenomenon of triaxiality has remained of central importance
to our understanding of galaxy dynamics since the demonstration 
by Schwarzschild (1979) of the existence of self-consistent triaxial 
equilibria.
Schwarzschild's models had large, constant-density cores
and the majority of the orbits were regular, respecting
three integrals of the motion.
But the realization that the central densities of elliptical
galaxies and bulges are very high (\cite{cra93}),
and that supermassive black holes are generic components of
galaxies (\cite{kor95}),
has modified somewhat our ideas about the persistence of triaxiality.
A central black hole can destroy triaxiality on large scales
by rendering the center-filling box orbits stochastic (\cite{geb85}).
Evolution to globally axisymmetric shapes can occur 
in just a few crossing times if the black hole contains
of order $10^{-2}$ of the galaxy's mass (\cite{meq98}; \cite{sel01}).

Less is known about the possibility of maintaining triaxial shapes 
at the very centers of galaxies, where the gravitational force is 
contributed roughly equally by the stars and by the nuclear black hole.
Regular orbits, both box-like and tube-like, exist in this region
(\cite{mev99}; \cite{sas00}; \cite{pom01}, Paper I) but the box-like orbits 
(the ``pyramids'') have shapes
that are generally contrary to the figure elongation, making them
less useful than classical box orbits for maintaining a triaxial shape.
Furthermore the pyramid orbits disappear within the ``zone of chaos,'' 
which begins at a radius where the enclosed stellar mass is a few
times that of the black hole (Paper I), roughly $10^2$ 
pc in a typical bright elliptical galaxy.

Here we demonstrate that triaxial equilibria do in fact
exist in the vicinity of the black hole.
We first (\S2) use Schwarzschild's technique to construct self-consistent
superpositions of orbits computed in a fixed triaxial potential
representing the stars and a central point mass.
We then (\S3) test the long-term stability of the models by using 
an $N$-body code to advance them forward in time; we find that the
$N$-body models maintain their triaxial shapes for many crossing times.
Finally we present some of the observable properties of the models
(\S4) and discuss the implications of triaxiality for observational
and theoretical studies of galactic nuclei (\S5).

\section {Schwarzschild Solutions}
We model the stellar nucleus as a triaxial spheroid with a power-law
dependence of density on radius,
\begin{mathletters}
\begin{eqnarray}
\rho_{\star} &=& \rho_{\circ} m^{-\gamma}, \\
m &=& \sqrt{\frac{x^2}{a^2}+\frac{y^2}{b^2}+\frac{z^2}{c^2}},\label{ellipr}
\end{eqnarray}
inside of some surface $m=m_{out}$ discussed below.
We took $\gamma=1$ or $2$; these values correspond roughly to the
density profiles observed at the centers of bright and
faint elliptical galaxies, respectively (\cite{geb96}).
\end{mathletters}
The triaxiality $T$ is defined as 
\begin{eqnarray}
T \equiv \frac{a^2-b^2}{a^2-c^2}.
\end{eqnarray}
We chose $a=1.0, b=0.79, c=0.5$ for both models, corresponding
to $T=0.5$, ``maximal'' triaxiality.
The central black hole is represented by a point mass with $\mh = 1$.
Expressions for the gravitational potential and forces corresponding
to this mass model are given in Paper I.

Although our mass model for the stars is scale-free, the presence
of the black hole imposes a scale.
We identify three characteristic radii associated with the black hole.
At a radius of $r_g$, the enclosed mass in stars 
(defined as the mass within an equidensity surface which
intersects the $x$-axis at $x=r_g$)
is equal to that of the black hole.
In real galaxies, $r_g$ is typically a factor of $\sim$ a few
greater than the ``sphere of influence'' $r_h$, where
\begin{equation}
r_h\equiv {G\mh\over\sigma_*^2} \approx 10.8\ {\rm pc} \left({\mh\over 10^8\msun}\right) \left({\sigma_*\over 200\ {\rm km\ s}^{-1}}\right)^{-2}
\end{equation}
depends on the stellar velocity dispersion $\sigma_*$
The third fiducial radius is $r_{ch}$, the radius at which
the character of the box-like orbits undergoes
a sudden transition to chaos in triaxial potentials.
Table 1 gives values of $r_g$ and $r_{ch}$ for our two mass models;
values of $r_{ch}$ are approximate and are taken from Paper I.
Table 1 also gives values of the dynamical times at $r_g$ and $r_{ch}$,
defined as the period of a circular orbit of the same energy in
the equivalent spherical potential, defined to have a scale length
$a_{ave}=(abc)^{1/3}$.
The enclosed stellar mass within $r_{ch}$ is about $3\mh$ for
$\gamma=1$ and $6\mh$ for $\gamma=2$.

\begin{tablehere}
\vspace{0.2cm}
\caption{\label{table1}}
\vspace{0.1cm}
\noindent\begin{tabular}{ccc}
\hline
$$ & $~~~~~\gamma=1~~~~~$ & $~~~~~\gamma=2~~~~~$ \\ \hline 
$r_g$ & $0.64$   & $0.20$  \\ 
$T_D(r_g)$ & $0.98$ & $0.17$ \\ 
$r_{ch}$ & $1.1$ & $1.3$ \\ 
$T_D(r_{ch})$ & $1.8$ & $1.5$  \\
$m_{out}$ & $5.6$ & $3.8$ \\ 
$T_D(m_{out})$ & $5.1$ & $4.6$ \\ \hline
\end{tabular}
\end{tablehere}
\bigskip\bigskip

Any numerical realization of a stellar system must have an outer boundary.
Our goal was to test self-consistency out to a radius of at least $r_{ch}$,
corresponding to $\sim 2 r_g$ for $\gamma=1$ and $\sim 6r_g$ for
$\gamma=2$.
We therefore chose the outer surface of our mass model, 
$m=m_{out}$, to be large enough that almost all of the density
at $r_{ch}$ in a real galaxy would be contributed by orbits with 
apocenters below this surface.
In order to estimate $m_{out}$,
the isotropic distribution functions corresponding to a spherical
galaxy with the density law of equation (1), $\gamma=\{1,2\}$, and a central
point mass were computed
and transformed to $F(r_{apo};r)$, the distribution of apocentric
radii $r_{apo}$ of orbits passing through $r$.
For $\gamma=1$, we found that orbits with $r_{apo}\lap 5 r_{ch}$
contribute $\sim 70\%$ of the density at $r_{ch}$.
For $\gamma=2$, setting $r_{apo}\approx 3 r_{ch}$ is sufficient
to give $\sim 75\%$ of the density at $r_{ch}$.
($r_{ch}$ in the spherical models was defined as the radius
containing the same mass as in the triaxial models.)
Our corresponding choices for $m_{out}$ are given in Table 1.

We followed standard procedures for constructing the Schwarzschild
solutions (\cite{sch93}; \cite{mef96}).
The mass model within $m_{out}$ was divided into 64 equidensity shells and each
shell was divided into 48 cells per octant, for a total of
3072 cells.
Shells were more closely spaced near the center (Figure 1).
Orbits were computed in two initial condition spaces:
stationary start space, which yields box-like orbits,
and $X-Z$ start space, which yields mostly tube orbits.
Orbital energies were selected from a grid of $42 (52)$
values for $\gamma=1 (2)$,
 defined as the energies of equipotential surfaces
that were spaced similarly in radius to the equidensity
shells.
The outermost energy shell intersected the $x$ axis at
$x=m_{out}$ for both models.
Orbits were integrated for $100$ dynamical times,
as defined above, and their contributions to the masses
in the cells were recorded.
In order to distinguish regular from stochastic trajectories,
the largest Liapunov exponent was computed for each orbit.
For $\gamma=1$, the total number of orbits integrated was
18144 of which 9751 were found to be regular.
For $\gamma=2$, the numbers were 22464 and 15048 respectively.
Orbital weights that reproduced the masses in the 3072 cells
were found via quadratic programming (e.g. \cite{dej89}).

\myputfigure{figure1.ps}{3.2}{0.5}{-10}{-5} 
\figcaption{
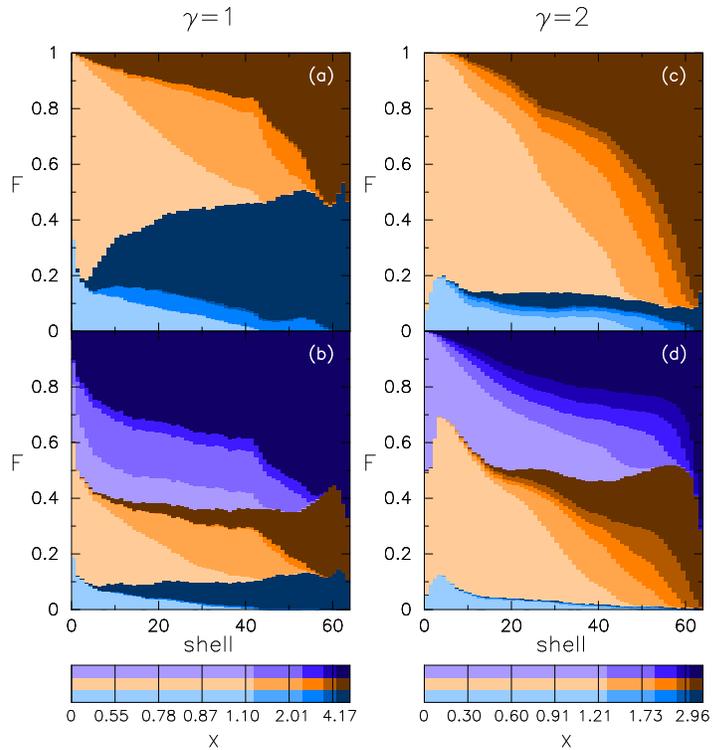
Cumulative mass fractions $F$ contributed by different 
kinds of orbits to different shells of the triaxial models.
Box-like orbits are green, tube orbits (both $z$- and $x$-tubes)
are orange, and chaotic orbits are purple.
Higher energies are indicated via darker shades; the color bar
relates the shade to the radius of the equipotential surface,
as explained in the text. 
Numbers below the color bar indicate radii where the equidensity shells
intersect the $x$-axis.
(a), (c) correspond to solutions with only regular orbits for
$\gamma=1$, $2$ respectively.
(b), (d) correspond to solutions with both regular and chaotic orbits.}
\vspace{\baselineskip}

As expected, orbital superpositions that exactly reproduced the
cell masses in all of the cells, including the outermost ones,
could not be found since only a few orbits visit the outer
shells.
We were able to find ``exact'' solutions (solutions which
precisely reproduced the densities in a subset of cells) by
relaxing the constraints in a number of ways; for instance,
by eliminating the angular constraints on the outermost shells
(forcing only the integrated shell mass to be fit),
or by ignoring the outer cells entirely when imposing the constraints.
For example, when only regular orbits were included, 
``exact'' solutions could be found for $\gamma=1$ for all cells 
within shell $51$, corresponding to a radius of $\sim 3.2$.
This radius is substantially greater than $r_{ch}$ (Table 1).

When these ``exact'' solutions were advanced forward in time, as described
below, they were found to exhibit significant evolution at large radii,
due to the fact that the model's density was not correctly reproduced
in the outermost cells.
We therefore constructed new solutions which were not ``exact,''
but which were constrained to reproduce the densities in {\it all} cells
within $m_{out}$ with as high an accuracy as possible.
Such solutions exhibited smaller fractional errors in the 
innermost cells ($r\lap r_{ch}$) than in the outer ones.
Models constructed in this way were found to evolve much less than
the ``exact'' solutions and provide the basis for the discussion below.
\begin{tablehere}
\vspace{0.2cm}
\caption{\label{table2}}
\vspace{0.1cm}
\noindent\begin{tabular}{ccccc}
\hline
$$ & $z$-tubes & $x$-tubes & pyramids & chaotic\\ \hline 
$\gamma=1 :$ (regular)& &  &\\
$r<0.5$ & 0.61 & 0.10 & 0.29 & 0.00 \\ 
$r<1.0$ & 0.56 & 0.08 & 0.37 & 0.00\\
$r<1.5$ & 0.53 & 0.07 & 0.40 & 0.00\\ \hline
$\gamma=1 :$ (all)~~~~&  &  &  &\\
 
$r<0.5$ & 0.09 & 0.04 & 0.26 & 0.61\\ 
$r<1.0$ & 0.10 & 0.03 & 0.25 & 0.62\\
$r<1.5$ & 0.10 & 0.02 & 0.25 & 0.63\\ \hline
$\gamma=2 :$ (regular) & & & &\\
$r<0.4$ & 0.75 & 0.10 & 0.15 & 0.00 \\
$r<0.8$ & 0.73 & 0.13 & 0.14 & 0.00 \\
$r<1.2$ & 0.73 & 0.13 & 0.13 & 0.00 \\ \hline
$\gamma=2 :$ (all)~~~~&  &  &\\
 
$r<0.4$ & 0.06& 0.05& 0.45& 0.45\\ 
$r<0.8$ & 0.05& 0.05& 0.44& 0.47\\
$r<1.2$ & 0.04& 0.05& 0.42& 0.49\\ \hline
\end{tabular}
\end{tablehere}
\bigskip\bigskip

The orbital content of the latter solutions is described in 
Figure 1 and Table 2.
We constructed two types of solution for each $\gamma$: 
solutions in which  only regular orbits were used; and
solutions in which all orbits (regular and chaotic) were provided
to the Schwarzschild algorithm.
The dominant family of orbits in the models containing 
only regular trajectories was found to be the $z$-tubes,
orbits which circulate about the short axis of the figure.
Plots of the individual orbits used in the solutions confirm that
many of the $z$-tubes have the correct shape for reproducing the
triaxial figure, being more elongated in $x$ (the long axis) than
in $y$ (the intermediate axis).
Most of the remaining contribution to the density 
of the regular nuclei was found to
come from pyramid orbits and high-energy box orbits 
in the $\gamma=1$ model, while for
$\gamma=2$, roughly equal contributions came from pyramid orbits
and from the $x$-tubes.
(See Paper I, Figure 1 for illustrations of the various orbit
families.)
The most heavily occupied pyramid orbits had shapes similar
to ``saucer'' orbits, $z$-tubes associated with a $2:1$ resonance
(Paper I, Figure 1c); both models contained roughly equal
numbers of saucers and pyramids.
High-energy box-like orbits are more important 
in the $\gamma=1$ case than in the $\gamma=2$ case, 
because: (1) high-energy tube orbits in the $\gamma=1$ model 
are elongated opposite to the figure;
(2) most of the high-energy regular box orbits in the 
$\gamma=2$ model are bananas, which are not very useful for filling the
outermost shells.

In the models constructed using both regular and chaotic trajectories, 
the number of stars on chaotic orbits is surprisingly high:
$\sim 60 \%$ for $\gamma=1$ and $\sim 45 \%$ for $\gamma=2$ at all radii.
Plots of the most highly populated chaotic orbits suggest that they have 
reached a nearly steady state in a time-averaged sense, filling the volume 
defined by the equipotential surface corresponding to their energy.
Stationary chaotic building blocks like these have been used before when
constructing triaxial models (e.g. \cite{mef96}) but no published triaxial
models have contained nearly so large fraction of stars on chaotic orbits.

\myputfigure{figure2.ps}{3.2}{0.45}{-10}{-5} 
\figcaption{\label{figure2}
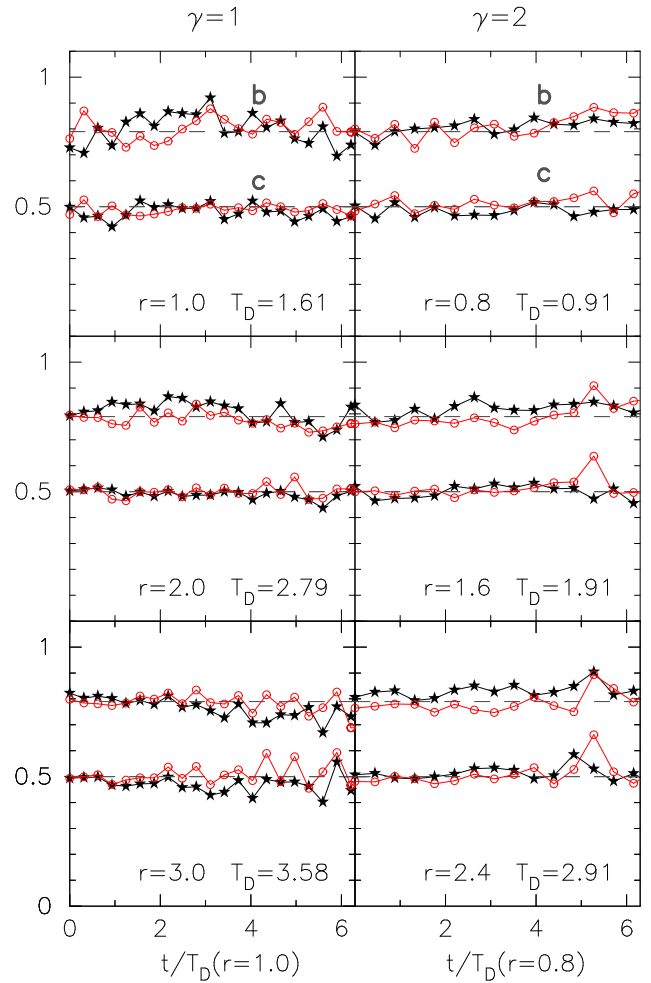
Evolution of the $N$-body axis ratios; $a\equiv 1$.
$r$ defines the longest dimension of the ellipsoid within which
the axis ratios were determined, using the algorithm described in the text.
$\mbox{T}_{\mbox{D}}$ is the dynamical time at this radius.
Black (red) curves correspond to solutions where
only regular (both regular and chaotic) orbits were used.}
\vspace{\baselineskip}

\section{N-body models}
We verified that our orbital superpositions correspond to true
equilibria by realizing them as $N$-body models and integrating
them forward in time in the gravitational potential computed from the
$N$ bodies themselves and from the point mass representing the black hole.
Initial conditions were prepared by re-integrating all orbits with
nonzero occupation numbers.
The position and velocity of each trajectory were recorded at
fixed time intervals; the length of each integration was determined by 
the orbit's occupation number.
The sense of rotation of the tube orbits was chosen randomly 
such that the mean motion was everywhere zero.
The number of particles was $2.42\times 10^6$ for $\gamma=1$ and
$2.19\times 10^5$ for $\gamma=2$; a smaller $N$ was chosen for
$\gamma=2$ since the $N$-body integrations were slower for the more
condensed model.

The initial conditions were then advanced in time using the 
$N$-body code {\tt GADGET} (\cite{syw01}), a parallel tree code
with variable time steps.
The interparticle softening length was chosen to be $0.005$ ($0.003$) 
for $\gamma=1$ ($2$).
Figure 2 shows the evolution of the axis ratios of the models,
computed using the iterative procedure described by 
Dubinski \& Carlberg (1991).
The axis ratios fluctuate with time,
but the models retain their triaxial figures very well over many
dynamical times.
No signficant evolution was seen in the radial distribution of matter.

\myputfigure{figure4.ps}{3.2}{0.55}{-10}{-5} 
\figcaption{\label{figure3}
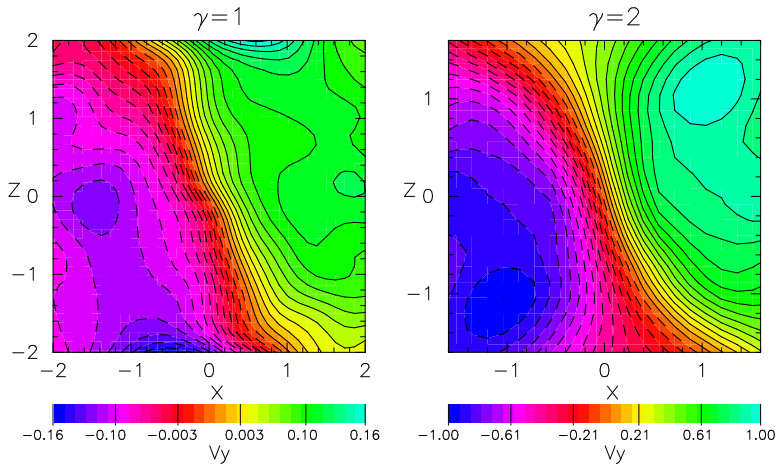
Line-of-sight velocities of the rotating regular models described
in the text, as seen along the $y$-axis.
Contours of positive (negative) velocity are represented by
full (dotted) lines.}
\vspace{\baselineskip}

\section{Observable properties}

Two observational signatures of triaxiality are
isophote twists, and kinematic misalignments.
Isophote twists require a variation with radius of the
intrinsic axis ratios, which is not a feature
of our models.
Kinematic misalignments occur whenever the intrinsic
angular momentum vector does not coincide with the minor
axis of the figure, and are generic features of triaxial
systems when both the short- and long-axis tube orbits are
populated (e.g. \cite{fiz91}).

We illustrate the possibility of kinematic misalignments in our
models in Figure 3.
Line-of-sight velocity contours are plotted for regular models
after a fraction of the $x$- and $z$-tubes were reversed,
giving the models net rotation about both the long and short
axes.
For $\gamma=1$, all x-tubes in Figure 3 rotate in the same sense while
60\% of the $z$-tubes at each energy rotate in one sense
and the remaining $40\%$ in the other.
For $\gamma=2$, all $x$-tubes rotate in the same sense around
the $x$ axis and all $z$-tubes rotate in the same sense around the
$z$-axis.
Rotation of the projected models along the apparent minor axis of the 
figure (``minor-axis rotation'') is apparent.

\section{Discussion}

We have shown that long-lived triaxial configurations are possible
for nuclei containing black holes, whether constructed purely from regular orbits,
or from a combination of regular and chaotic orbits.
The latter solutions are remarkable given the large fraction of the mass
on chaotic orbits, of order 50\% or more (Table 2).
Such solutions violate Jeans's theorem in its standard
form (e.g. \cite{bt87}) but are consistent with a generalized
Jeans's theorem (\cite{mer99}) if we assume that the chaotic
building blocks are ``fully mixed,'' that is, that they approximate
a uniform population of the accessible phase space.
This appears to be the case based on visual inspection of
the time-averaged chaotic trajectories.
While a sudden onset of chaos can effectively destroy triaxiality in 
models containing a large population of regular box orbits
(\cite{meq98}; \cite{sel01}),
our work shows that at least the central parts of galaxies 
containing black holes can remain triaxial even when dominated 
by chaotic orbits.

Galaxy nuclei are typically modelled as oblate axisymmetric systems when 
deriving black hole masses from stellar-kinematical data 
(e.g. \cite{vdm98}; \cite{jos01}).
Modelling a triaxial nucleus as axisymmetric will generally lead to 
biassed estimates of $M_{bh}$; for instance, if an elongated bar is
viewed end-on, the velocity field can mimic that produced by a black
hole even in the absence of a central mass concentration (\cite{ger88}).
The high-resolution data on which such mass estimates are 
based are usually obtained from single slits, making it difficult to rule out
triaxial shapes using information like that in Figure 3.

Axisymmetry is typically also assumed when deriving rates of interaction of
stars with nuclear black holes (e.g. \cite{syu99}; \cite{mat99}).
The triaxial models which we constructed from regular orbits were  
dominated by $z$-tubes (Table 2), similar to the tube orbits that populate 
an oblate spheroid.
The rate of capture or destruction of stars by the black hole in these 
models would
be similar to the rates in an axisymmetric nucleus.
However the dominance of chaotic orbits in our second set of solutions
implies a qualitatively different sort of behavior: the ``loss cone'' of stars
captured by the black hole would be refilled on a time scale determined
by the frequency of near-center passages of the chaotic orbits, 
rather than by two-body scattering.

This work was supported by NSF grants AST 96-17088 and 00-71099 and
by NASA grants NAG5-6037 and NAG5-9046 to DM.

\end {document}